# Comparing Field Trips, VR Experiences and Video Representations on Spatial Layout Learning in Complex Buildings

Cetin Tuker [a*], Togan Tong [b]

[a] Mimar Sinan Fine Arts University, Graphic Design Department, Fındıklı Kampüsü, Meclisi Mebusan Cad., Beyoğlu, İstanbul, Turkey

[b] Yıldız Technical University, Department of Architecture, Yıldız Kampüsü, Barbaros Bulvarı., Beşiktaş, İstanbul, Turkey

Corresponding Author: Cetin Tuker*
Mailing Adress:
Mimar Sinan Fine Arts University,
Graphic Design Department,
Fındıklı Kampüsü,
Meclisi Mebusan Cad.,
Beyoğlu, İstanbul, Turkey
cetin.tuker@msgsu.edu.tr

# Comparing Field Trips, VR Experiences and Video Representations on Spatial Layout Learning in Complex Buildings

## Abstract
This study aimed to compare and investigate the efficacy of the real-world experiences, immersive virtual reality (IVR) experiences, and video walkthrough representations on layout-learning in a complex building. A quasi-experimental, intervention, and delayed post-test research design was used among three groups: real-world, IVR, and video walkthrough representation. A total of 41 first-year design students from architecture, and game design departments were attended the study. Design students were selected as they already know how to communicate graphically the layout of a building on paper. IVR and video walkthrough groups experienced the representations of the building by themselves, but real-world group experienced the building within a group as imitating a design education field trip. After 10 days, a total of 26 participants out of 41 tested for spatial recall performance. Recall performances were measured as an indicator of layout learning by analysing the plan sketches drawn by the participants. Results showed IVR group recalled significantly more spatial memories compared with the real-world and video walkthrough representation groups. Real-world group performed the worst among three groups. This result was interpreted to three conclusions. First, the layout learning tends to be overly sensitive to distractions and lower levels of distraction may lead to better levels of layout learning. Second, for the domain of layout knowledge learning, a remarkably simple IVR model is enough. Third, although real-world experiences give direct and richer sensory information about the spaces, layout knowledge acquisition from the surrounding environment requires psychologically active wayfinding decisions.

# 1. Introduction

People acquire spatial knowledge from two types of sources: (a) direct sources, the space itself which was experienced by the people directly by sensory-motor experiences (sitting, standing, walking, moving, in the space); (b) indirect sources, which are the external representations (plans, elevations, sections, maps, video walkthroughs) of the space (Montello, Waller, Hegarty, Richardson, 2004, p.252). In the former, the scale of the space is larger than the observer and the observer is contained by the space (in other words, space surrounds the observer) and in the latter, the scale of the representation is smaller than the observer, and it is observed from outside of the representation (Montello et al. 2004, p.252). When space contains the observer, the spatial data from the environment is direct, and the observer can create the mental image of the space easily.

Indirect sources are pictorial representations like plans, elevations, and sections and use abstract symbols to represent spatial information. Symbolic representations and verbal narratives of space contain insufficient data compared to direct on-site experiences due to limitations of the media used in the presentation process. They need more cognitive load compared to direct sources because of the mental processes needed to convert or decipher the symbols to a meaningful mental image of the space (Montello, et. Al., 2004, p.255).

As the level of mental visualization abilities of people is not the same, the correctness of the created mental representation of the space depends on the spatial skills of the individual. Previous studies show that there are differences between the experience of the three-dimensional environment directly on-site and visualization in mind through pictorial representations in terms of the amount of information transmitted from human sensory organs and the brain regions in which the information is processed (Hegarty, Montello, Richardson, Ishikawa, Lovelace, 2006)(Table 1).

Table 1
Information Types of Various Experiences

|  | **Sensory-Motor Information** | **Vision** | **Data Source** |
| --- | --- | --- | --- |
| **Direct Experience** | Eyes, locomotion, vestibular system | Stereoscopic | Direct (little need of mental visualization) |
| **Video Walkthrough** | Only eyes | On a 2D flat screen | Indirect, series of images (Requires mental visualization) |
| **IVR Experience** | Limited tracking of HMD. Limited movement of viewer | Stereoscopic but field of view is not equal to natural vision | Direct (little need of mental visualization) |

In all cases, the viewer needs some level of cognitive power to create allocentric interrelations between the locations of landmarks and the initial layout to create route knowledge (the knowledge of direct connections, walk paths from one space to the other) between the landmarks. While receiving visual data, any distraction (environmental factors, fidelity issues of interfaces, human factors) will result to a decrease in the level of spatial data stored in the brain.

Existing literature mostly focuses on how users of the space create cognitive maps of the space in the navigation context. Three types of spatial learning types emerge in literature: (a) route knowledge, recalling the relations of routes or corridors in a built environment that provide navigation without getting lost; (b) survey knowledge, correctly orienting oneself in the environment and assuming distances between parts of the environment; (c) landmark knowledge, remembering the landmarks in the environment (Hirtle, 2009).

On the one hand, it can be claimed that direct on-site experiences can provide more spatial information and spatial experience (Hegarty, et. Al., 2006) which can lead to a richer spatial learning outcome, on the other hand, it is not logistically, and economically viable and practical, and can be very time-consuming activity to experience real-world environments in some situations. Educational settings related with spatial learning (like architecture and design) can be major examples of these kinds of situations. Instructors mostly prefer to teach spatial relations by using indirect sources, as it is more practical to prepare educational material, and only very little part of the courses can be reserved for direct on-site experiences due to the logistical shortcomings mentioned above. Even if the student can create a mental image of the certain space by using indirect sources, it is open to question whether the student could successfully create the mental image of the space or benefit from it to contribute to education. Researches show that mental visualization skills are not equal for individuals and (Cherney, 2008; Terlecki & Newcombe, 2005; Uttal et. Al., 2013), and lack of mental visualization skills can increase the probability of dropping out of university in engineering education (Sorby, 2009). Additionally, group interactions such as a crowded group that interacts with each other instead of observing the spatial interrelations during on-site experiences can demand the attention of the student from the educational subject, and this could result in a decrease in the amount of spatial learning.

Within this context, stereoscopic immersive virtual reality (IVR) experiences can be an intermediate level platform between direct and indirect experiences. Immersion is a technology (mostly hardware) that simulates the real-world equivalent of a sensory input modality (Slater, 2003). According to Berkman & Akan (2019), "immersion is the user's engagement with a VR system and depends on sensory immersion" (p. 1), which is provided by the equipment. IVR technology is not new, and as technology develops, IVR systems are getting more available for educational institutions. Although IVR experiences in the future may be an alternative to direct experiences, it must be investigated by experiment that whether the idea of using IVR systems in the instruction of spatial knowledge can be supported with measurable and replicable scientific evidence. The reasons how, why, and in what limits students' learning the spatial knowledge from IVR experiences, is another subject to investigate. Therefore, this study aims to compare the real-world experiences (field trips), IVR experiences, and video representations on spatial learning. To accomplish this goal an experimental research study was conducted by measuring the spatial recall performances of participants among three conditions: (a) real-world spatial experience which is also the control group; (b) video recorded spatial experience; (c) virtual reality experience with tracked head-mounted display.

## 2. Background
### 2.1 How fast the spatial knowledge is formed?
One of the discussions in the area is about how fast spatial knowledge is formed. In one of the most prominent research, Moeser (1988) shows survey knowledge does not form automatically in student nurses working in a complex hospital building even after three years of experience. In the study, experienced student nurses failed to draw maps (schematic) of the hospital they were working. A naïve control group that studied the same building from a map performed better in room-labeling, direction, and distance estimation tasks compared with student nurses. The study of Rossano, West, Robertson, Wayne, & Chase, (1999) supports these findings in survey type learning. In this study, there were three groups. A computer model walkthrough was used to train participants. Another group studied the building by using a map. The last group was the users of the building. In route tasks, the real-world group was superior to others. But in survey knowledge tasks, map group were superior to other groups. It was reported that the real-world group and computer group, needs time and repeated exposures to learn the spatial relations. The study of Waller, Hunt, & Knapp (1998) supports these findings as long exposures with the surrounding environment (IVR) affect spatial learning in a positive way. The study of Thorndyke, & Hayes-Roth's (1982) findings contradicts with Moeser (1988) for females. In this study, straight-line distance, root distance, and direction estimation, and route-finding task, performances of naïve users that studied the building from maps were compared with the users of the building. With moderate exposure, naïve users were superior to users of the building that were using the building for 1-2 months. With extensive exposure (6-8 months and 10-11 months), this superiority vanishes, and the real-world experience becomes superior for route distance and orientation tasks (Thorndyke, & Hayes-Roth, 1982). These findings show different methods can be used for different learning objectives and possible effects of other variables such as gender difference, complexity or landmark layout of the building plan, or wayfinding strategies of the users in addition to exposure duration.

### 2.2 Spatial Knowledge Training with Media and Types of VR Representations
Another discussion is about training types and efficiency of media and methods used in spatial knowledge training. There are two learning modes exists in the literature concerned with spatial learning in virtual environments. These are physical and psychological modes and both have active and passive conditions (Carassa, Geminiani, Morganti, & Varotto, 2002; Wallet, Sauzéon, Rodrigues, & N'Kaoua, 2008; Wilson, Foreman, Gillett, & Stanton, 1997). Phsically passive condition means participant experiences the virtual environment without a body movement (like watching a video) and being active means participant controls the movement by using a joystick or keyboard like device. Psychologically passive means participant is being guided by an avatar or the movement within the environment is restricted, whereas psychologically active is exploring the virtual environment freely .

Comparing participants' spatial learning with walkthrough training (that participants passively watch the camera traverse the environment without interaction) with a map, pictorial presentations, and real-life experience is a frequently used study design. With the development of virtual reality (VR), researches include training with virtual environments with different capabilities. Broadly there are two types of VR systems mentioned in literature: (a) three-dimensional models and environments viewed on two dimensional flat monitors (these can be classified as indirect pictorial representations -VR, desktop virtual environment); and (b) three-dimensional environments and models viewed with stereoscopic head-mounted displays (HMD) and where both HMD and the locomotion of the user can be tracked (immersive VR - IVR). The former is in the first-person perspective and either actively controlled by the user or passive walkthrough animations that the user cannot interact with the motion of the camera. The latter, IVR, can be either in the form of a stationary user (mostly sitting) or the user can move in the space with a tracked HMD with 6 degrees of freedom. With the development of computer technology, new IVR systems have more smooth tracking systems and are easy to use for research.

The former is closer to indirect experiences, and data collected from that kind of representation is stored in human memory as a series of images that needs to be mentally constructed to a mental representation of the space that was experienced. The latter, with its ability of locomotion tracking and providing stereoscopic images that creates the feeling of the viewer, that is surrounded by the environment (immersive VR - IVR) can simulate the immersion and can be an alternative to direct experiences for spatial learning.

### 2.3 Effects of Spatial Abilities and Media on Spatial Learning
The relations between spatial abilities and spatial learning are also subject to investigation. In the study of Hegarty, Montello, Richardson, Ishikawa, & Lovelace (2006), participants were trained with real-world experience, video walkthrough, and interactive desktop VR. The study draws connections with the scale of the environment focusing on the dissociation model of participants' learning of spatial layout of a building either from small scale (viewed on a computer screen) or large scale (environment that surrounds the viewer like direct experience) presentations of spaces. 221 participants were tested on psychometric measures of spatial and verbal abilities and working memory. In this study, small scale presentations were a walkthrough video and a first-person view (like a video game) interactive walkthrough VR experience of the same building, and large-scale presentation was the direct experience (real-world) guided by a researcher in the corridors of the same building. Findings show that learning from small scale presentations is related to the small-scale visualization ability of the participant. In other words, not all participants can learn spatial layouts from video and game-like VR presentations represented on flat computer screens (Hegarty et. Al., 2006).

Ruddle, Payne, & Jones (1997) and Albert, Rensink, & Beusmans (1999) used only desktop VE's to train and assess participants. In the former study, landmarks in VE's affects the wayfinding performances of participants positively (Ruddle et al.,1997). The latter study supports these findings that desktop VE creates viewpoint dependent representational memories Albert et. Al. (1999). To compare wayfinding performances of participants trained by using an animated walkthrough with pictorial presentations of decision points in a building, the animated walkthrough group performed better than the pictorial presentation group (Münzer, & Stahl; 2011). In other words, participants store images of the landmarks instead of creating a layout map of the building in their mind and locate the landmark on that plan and linking the connection between the landmarks with a series of images of the route help perform better. Moeser (1988) also shows student nurses cannot draw maps of the hospital that they were working for almost three years and use landmarks to find their way in the building. These studies show creating maps of an environment is not an automatic process, or it is easier to store route information with a series of viewpoint dependent images.

Interactive game-like VE's also affects spatial learning. Tüzün, & Özdinç, (2016) compared route learning performances of participants trained in a multiuser online VE with participants who were trained by a researcher on a map while learning the spatial layout. In this study, the VE group recalled route details better than real-world group (which was guided by the researcher during spatial learning session). The study of Farrell, Arnold, Pettifer, Adams, Graham, & MacManamon (2003) supports that users that make active navigational decisions learn more than the users guided by the researcher that already knows the route. Waller et. Al. (1998) also supports these findings. Waller et al. compare six different groups that learn the plan of a maze from no-training, real-world, map, desktop VE walkthrough, IVR short-exposure, IVR long-exposure. They were assessed with a wayfinding task in a real-world maze. In this study with long exposure training, the IVR group surpasses real-world group in the wayfinding task. This shows to acquire route knowledge in IVR training time, repeated exposures to training material, and active navigational decisions are essential. Active learning has an impact on the level of spatial learning. Clawson, Miller, Knott & Sebrechts, (1998) studied the effects of user control on indoor navigational learning. Participants were trained with no-user-controlled VR, limited-user-control VR, full-user-controlled VR. Results show, as user control of VR increases, learning level increases in route learning tasks. In this study, the full user-controlled VR group has the maximum level of route learning. Findings of Farrell et. Al. (2003) supports Clawson et. Al.'s (1998) findings. In a study that aimed to investigate transfer of route knowledge in desktop VR to real environment, three training regimes were followed; train on map; active VR; and passive VR. Groups were tested on real life. Findings show that active VR with map training produces more transfer than passive VR or no VR training, and map with passive VR is not superior to active map only. Writers conclude active navigational decisions create spatial learning.

Some researchers claim training spatial knowledge with IVR systems also bring issues emerging from the field of view (FOV) of the virtual camera, the HMD, fidelity issues, and cognitive workload because of the interface. In one of the experimental studies (Henry and Furness, 1993) focuses on the differences of perception of real-world and virtual-spaces, the participants (architects, n=24) were asked to assume spatial dimensions and their orientation with respect to spatial landmarks while experiencing the same space in four different conditions that are direct experience, VR (non-immersive) on a 2D monitor, immersive VR with HMD tracking and immersive VR without HMD tracking. It was reported that findings favor the real experience over VR experiences as the subjects experienced the VR (in all three groups) guessed the dimensions of the spaces significantly smaller (this effect was maximized in immersive VR with HMD tracking) than the subjects experienced the real environment. Writers conclude this issue is probably because the FOV of the virtual camera used in the VR experiences was wider than the FOV of the human eye. Findings of Miller, Clawson, & Sebrechts (1999) partially support this finding. Participants that learned the plans and landmarks of an existing building from maps, trained with the actual building, and trained with IVR with HMD (78 degrees FOV) assessed with route finding and direction estimation tasks in the real building after two weeks both in aligned and reversed direction. IVR group was better only in route tasks with the same orientation with the training material but less accurate in survey tasks (in IVR), possibly because of low fidelity interface or excessive cognitive workload during training. For real-life tasks, although the IVR group is superior to the map group, training in real building influenced route learning the most. Contrarily, Findings of Johnson & Stewart (1999) does not support the claim that narrower FOV has a negative effect on spatial knowledge acquisition. A study made on soldiers who experience an IVR environment by using widescreen HMD's, small screen HMD's, and rear projection widescreen displays shows no difference in spatial knowledge, although there is a difference in perceived presence among HMD's and widescreen display.

## 2.4 Spatial Learning in Architectural Education
Field trips are widely used in the context of architectural education to help students to gain spatial experience from existing and outstanding buildings. As it is not logistically efficient to organize field trips every time for a large architectural class, the photographs, plans and sections of the buildings are also studied in architectural studios as a training material. However, the differences in how spatial learning occurs from field trips and from various media (representational tools) in the context of architectural education is not a well-studied area. There is also a very little amount of studies exist that try to create a link between human cognitive research and architectural design. In one of these studies, the wayfinding strategies of experienced and inexperienced users in a complex multi-level building were investigated (Hölscher, Meilinger, Vrachliotis, Brösamle, & Knauff, 2006) and architectural elements, mainly staircases that can possibly cause navigation problems were identified. About wayfinding, Lawton (1996) studies the self-reported strategies in indoor wayfinding and compares the results with the outdoor wayfinding strategies. In the same study, writers investigate the relationships of wayfinding strategies with spatial anxiety levels, and gender was also studied. Memikoğlu (Memikoğlu, İ. S. 2012. Exploring vertical navigation within a virtual environment: a staircase experience. Unpublished doctoral dissertation, Bilkent University, Ankara, Turkey) focuses on the vertical navigation problem targeting the process of the architectural design of multi-level buildings by articulating the problem by adding the parameter of the angle between staircases with respect to the location of the user. In the same study, staircase preferences of different genders in a virtual environment were studied in order to draw conclusions that can be used to guide designers with the design process.

## 2.5 Method and Media Debate in Instructional Technology
The last discussion at this point is the essential debate of method and media. According to Clark (1994), media is just a vehicle that delivers the instruction to be learned to learners, and it will not affect the achievement of learning goals. The main factor that influences learning is the instructional method. In this study, video and IVR are different types of media. Media can have several attributes like zooming in and out to details or slow-down the video-play to help a detail to be seen better, but these are not the main structural element of learning. As supported by evidence, the same learning goals can be achieved by different attributes of various media (Clark, 1994, p.22). The structural element of learning is the instructional method chosen for the specific need of the instruction (Clark, 1994). As an example, the learning-by-doing methodology is a standard method being used in design education. Focusing on the cost-effectiveness of several media, Clark claims, it is better to choose the media that leads to a cheaper, more comfortable, and faster way to deliver the same instruction (1994). In this study, as IVR simulates the real-world environment, theoretically, learning influenced by IVR cannot be larger than real-world experience. However, considering the other dynamics, in practice, IVR can have some positive attributes that can influence spatial layout learning.

**2.6 Hypothesis**
In this study, the hypothesis is related with the recall performance of the students in a complex building and claims that the students that experienced real-world field trip will be able to recall more spatial memories than VRE and video walkthrough experiences whereas the recall performance of the VRE group will be higher than the video walkthrough group.

**3. Method**
This study addresses the problem in measuring spatial recall (remembering) performance of the participants after seven days to ten days, as being able to recall information is one of the necessities of learning. This is a quasi-experimental, spatial experience, and delayed post-test research design.

**3.1. Research Context and Participants**
Participants of the study were the first and second-year video game design and architecture students of two different universities, and according to educational settings, it was not possible to select participants randomly. Therefore, participants of the same experiment group come from the same instructors' class. Researchers have no background information about how these students grouped into classes of different instructors.

A total of 41 participants attended the spatial experience part of the study. None of them has ever been in the Mimar Sinan Fine Arts University (MSFAU) building before. 30 of them were from the Architecture Department of Yıldız Technical University (YTU), and 11 of them were from Bahçeşehir University Game Design Department (BAU). All participants were first-year students, and all of them had just started (3 weeks to 6 weeks) their first semester at the date of the study. The male-female ratio of the groups was almost half. The total number of participants in groups are, 14 students for the real-world experience group; for the video experience group, 11 students; and for VRE group, 16 students. A total of 26 students out of 41 attended the second part of the study. Distribution of the participants by experiment groups were 7 students for real-world group; 8 students for video experience group and 11 students for IVR experience group (Table 2).

Table 2
Distribution of Participants According to Their Universities

|  | 1st Part | 2nd Part |
| --- | --- | --- |
| Real World (Control) | 14 (Architecture Students) | 7 |
| Video Walkthrough | 11 (Architecture Students) | 8 |
| Virtual Reality Environment | 16 (5 Arch. – 11 Game Design Students) | 11 (5 Arch – 6 Game Design) |
| Total | 41 | 26 |

**3.2 Preparation of the study environment**
MSFAU Architecture Faculty building (Fig. 2) was selected as the study environment. A walkthrough path that covers almost all corridors and three main stairs of the building were selected by the researchers. The researchers video-recorded the selected walkthrough path with an iPhone 7 Plus. The field of view (FOV) angle of an iPhone 7 Plus is 73.5 degrees (Fig. 1, Top). The duration of the recording was 11 minutes, and the camera was facing a strictly forward direction (first-person view) during the video recording. The whole walkthrough was recorded in a single shot. There was no sound channel in the video.

For IVR experience, the three-dimensional model of the building was created by using SketchUp software. Level of detail of the model was very low. Although all the architectural features of the building were modelled, most of the surface textures, details like ceiling beams, hanging lamps and sculptures were not added to the model. By using this model, a three-dimensional experience was programmed in the Unity game engine by the researchers (Fig 1, Bottom), which uses an HTC-VIVE IVR system as the visualisation hardware. The head mounted display of HTC-VIVE has 6DOF locomotion capabilities in 3 by 3 meters range and can be tracked by the two external tracking stations. So, users can freely move and look around within this range. Users interact with the IVR environment by using the two hand controllers. The horizontal FOV of the HTC VIVE is 110 degrees, whereas the peripheral vision of the human eye is 60 degrees for near peripheral, 120 degrees for mid-peripheral, and 220 degrees, including the far peripheral area (Adithya, Kumar, Lee, Kim, Moon, & Chai, 2018).

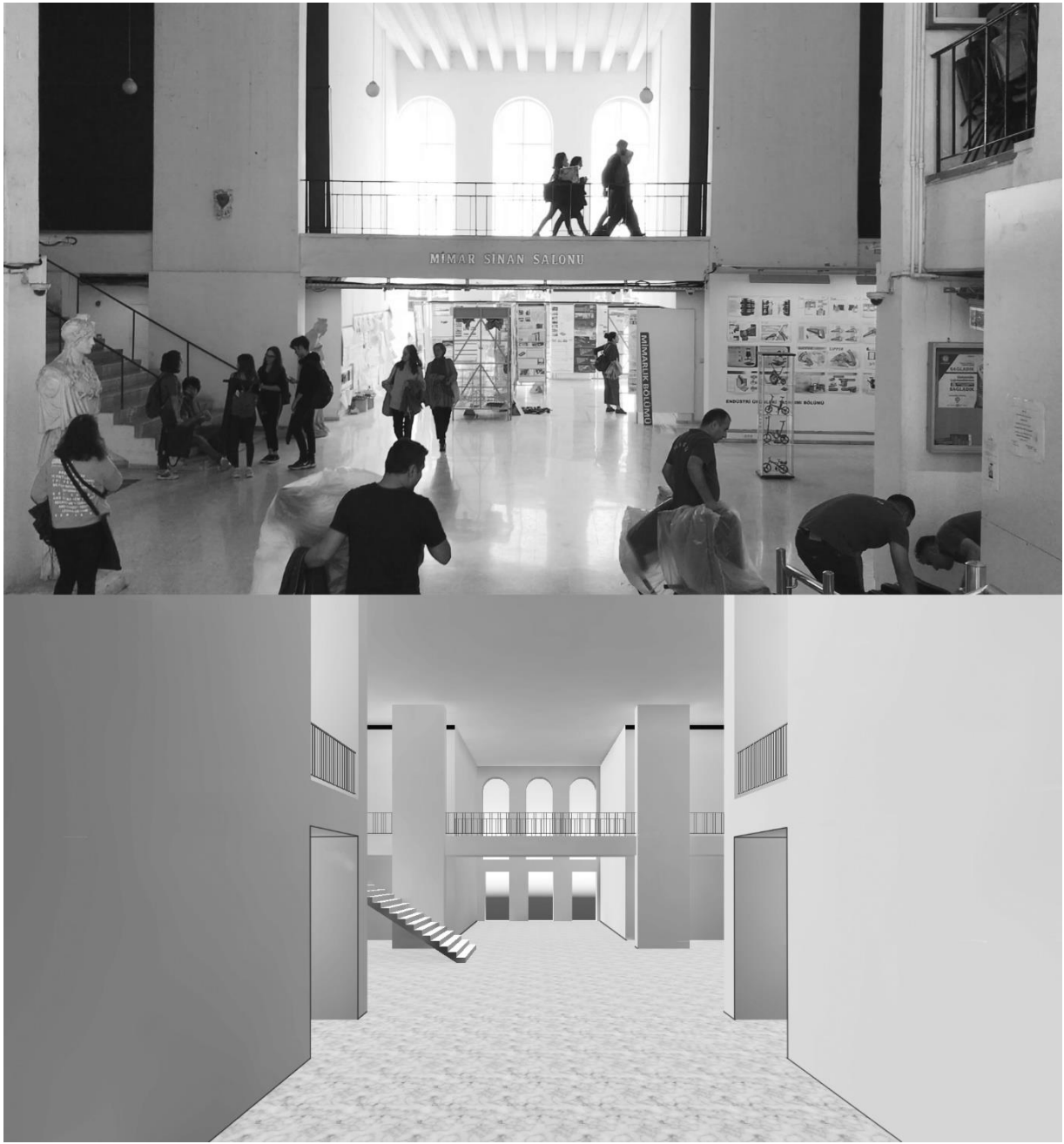

Fig 1
Top: First frame of video looking from the main entrance to main atrium of MSFSU Architecture faculty.
Bottom: Starting point of VR experience looking from the main entrance to main atrium of MSFAU Architecture Faculty.

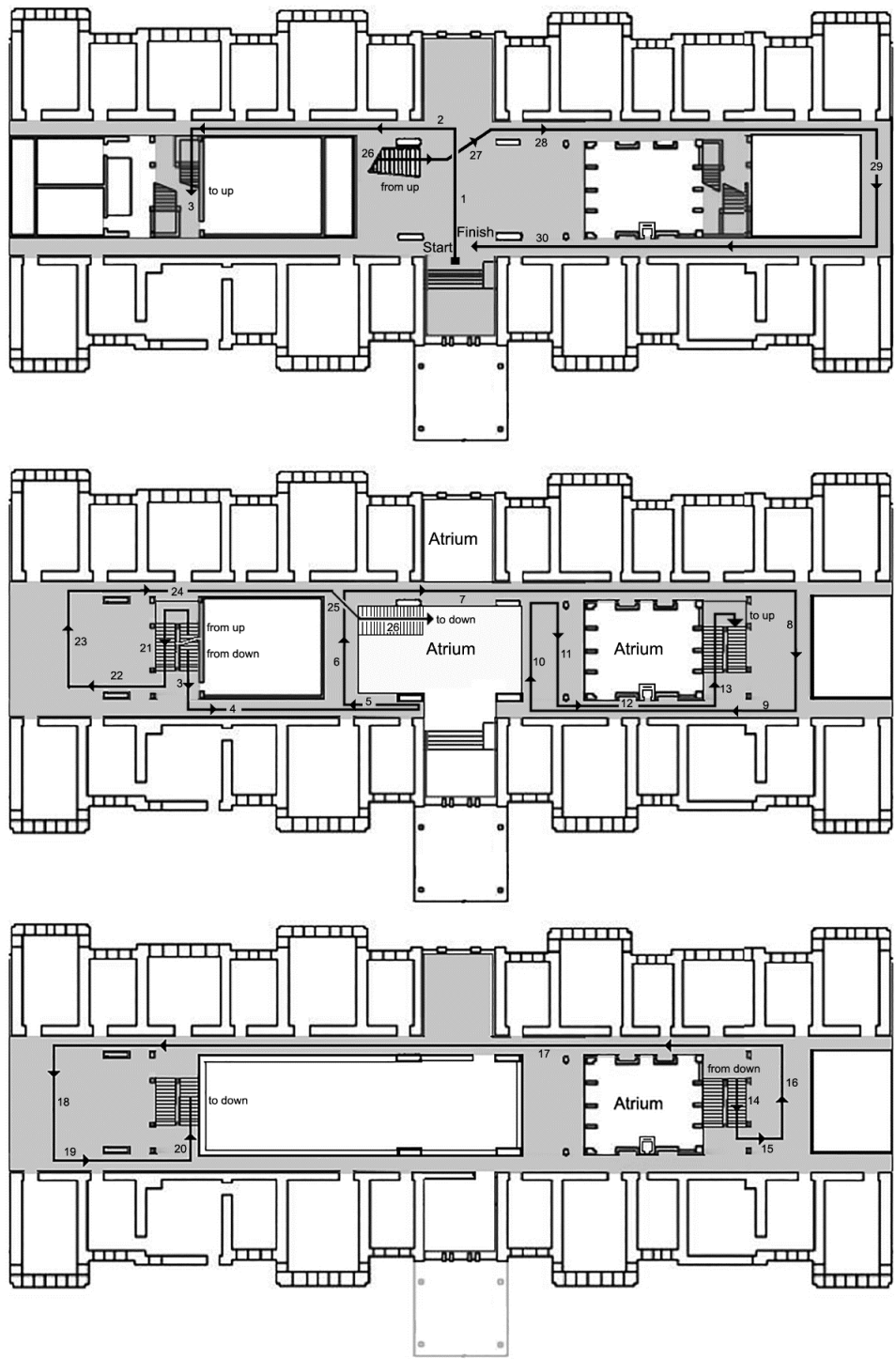

Fig 2
Floor plans of MSFAU Architecture Faculty Building. Arrows indicate the route that participants followed. Numbers indicate route segments.
Top: Entrance floor. Middle: First Floor. Bottom: Second Floor

### 3.3 Procedure
Researchers contacted with the participants by the help of their course instructors. The aim of the research was explained clearly to the instructors first. Next, researchers and instructors scheduled meetings with the student groups. Before or during the experiences, no instructions were given related with the aim of the research, about the building or any other subject to students to avoid a possible interference between the instructed material and spatial learning. Participants didn't know any information about the spatial relations and plan layout of the MSFAU building which was selected for the experiment also. Participants didn't know why they were experiencing the representations during the experiences.

In order to measure the memory recall performance of participants, a recall test was administered seven to ten days after the experiences. All participants have been contacted by email and they were asked to recall spatial memories of the building they experienced and to sketch a basic scheme or layout of all the floors they walked on. While sketching the layout, they were asked to locate their starting point on the walkthrough path, locate atrium openings, and vertical circulation shafts (staircases and elevators). 26 participants out of 41 replied and sent sketches (7 for the real-world group, 8 for the video group, 11 for the IVR environment group). (Fig. 3).

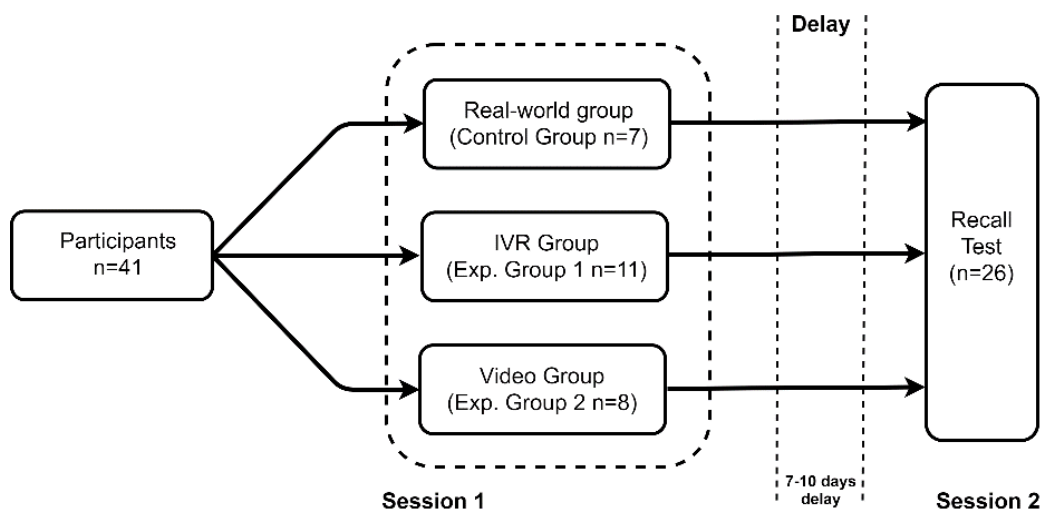

Fig 3
Schematic diagram of research study design

### 3.3.1 The IVR Group (Experimental Group 1)
Participants of the IVR group attended the experience one-by-one as there was only one HTC-VIVE HMD exist. Researchers briefly instructed them how to use the HMD's and the hand controllers. Although HTC-VIVE can track the user while moving in a 3x3 meters area, in this research, the main technique of moving from one point to another was teleporting in the IVR environment. Teleporting in IVR means the used is being transported from one teleport point to another teleport point without walking the distance between them. This was the only way they can move in a building of that size. The consecutive teleport points which have almost 3-5 meters in between were located on the same path that other experiment and control group followed during the experience. The IVR environment was designed in such a way that participants can follow the same walkthrough path in the same duration (11 minutes) as the other groups.

### 3.3.2 The Video Group (Experimental Group 2)
Participants of the video experience group watched the video in the computer laboratory at the same session on desktop computers (11 minutes). The screen size of the monitors was 21.5 inches and they were LCD flat monitors. All the participants remained seated during the video experience.

### 3.3.3. The Control Group
The researchers scheduled a meeting with the instructor of the real-world group in front of the main entrance of the MSFAU building. Participants were instructed briefly that how the experiment will be conducted. The participants of the real-world group attended the experience at the same time. During the experiment, the group walked the exact same path in the same duration (11 minutes). One of the researchers guided the group walking before them.

### 3.4 Data collection
Seven to 10 days after the experience, participants have been contacted by the researchers by email, and asked to recall their memories to make a sketch of the layout of the building. Layout sketching method was used in this study to measure how many of the architectural features of the building was recalled by the participants. Layout sketching method was used by Moeser (1988) by sketching the spaces (rooms) by the participants on a paper and labelling them. We improved the Moeser's method to a more quantitative method to make it free from the sketching skills of the participants. Instead of sketching the spaces of the building and labelling them correctly researchers asked participants to locate the key circulatory features (staircases, elevators and atriums). The sketches of the participants were graded by counting all features they located correctly on the layout sketch with respect to their starting point of the walkthrough. The upper limit of the grade a participant can achieve was 30. There are 14 features that exist on the original plan of the building (7 staircases, 3 elevators, 4 atrium openings. Fig. 4). If the participant recalls any of the features and sketches it on the paper (existence), it is graded by 1 point. Additionally, if the location of the same feature with respect to the starting point is correct, they get an additional 1 point (like on the left/right of the start point). The maximum points can be scored form features is 28. Additional one point comes from locating the entrance (start point of the experience) of the building correctly, and the last point is scored if the overall geometry of the plan sketched on the paper by the participant looks similar to the original plan (which is an elongated rectangle). If they are not similar enough, the participant can get 0 to 0.5 points. For every missing feature, the participant gets 1 missing feature point to be added to the missing feature score. Finally, if the participant remembers the feature but places on a different place than it should be, this was graded as 1 point for the feature score and 1 point for misplacement score. If the participant places more features on the sketch than should exist on the plan, he/she gets 1 misplacement point for every unnecessary feature.

### 3.5 Data Analysis
To calculate if at least one of the differences between the mean values of the three independent groups is statistically significant we conducted a one-way-ANOVA analysis. Due to the differences in group sizes we conducted a Scheffe's test as the post-hoc comparison test since it does not consider the assumption that the number of observations in groups is equal.

### 4. Findings
There was a significant effect of media on recall performance at the p<.05 level for the three conditions [$F(2,23) = 7.43$, $p = 0.002$]. Post-hoc comparisons using the Scheffe test indicated that, the mean score for the IVR experience ($M = 16.68$, $SD = 5.61$) was significantly different ($p=0.024$) than the video experience ($M = 9.5$, $SD = 3.30$) and the direct experience ($M = 7.43$, $SD = 6.25$, $p=0,005$). There is no significant difference between the recall scores of the video experience and direct experience groups. It can be concluded that IVR experience group remembers recall more spatial layout data compared with the other two groups (Table 3).

Missing feature performances of the groups were also investigated. The effect of media on missing feature performances was significant at the p<.05 level for the three conditions [$F(2,23) = 7.17$, $p = 0.004$]. Post hoc comparisons using the Scheffe test indicated that, the mean score for the IVR experience ($M = 6.09$, $SD = 2.77$) was significantly less ($p=0.005$) than the real-world experience ($M = 10.86$, $SD = 3.13$). There was no significant difference between the IVR experience and the video experience ($M = 9.00$, $SD = 2.07$, $p=0.087$). There was no significant difference between the missing feature scores of the video experience and direct experience groups. It can be concluded that IVR experience group remembers recall more spatial layout data compared with the other two groups by means of missing features (Table 4).

There was no significant effect of media on recall performance by means of misplaced features at the p<.05 level for the three conditions [$F(2,23)=0.931$, $p=0.408$] (Table 5). All mean scores of groups are given in Table 6. Some Examples of the sketches of recall task are given in Fig. 5.

Table 3
Spatial Recall Performance of Features Placed Correctly ANOVA Table

|  | Sum of Squares | df | Mean Square | F | Sig. |
| --- | --- | --- | --- | --- | --- |
| Between Groups | 437.390 | 2 | 218.695 | 8.043 | 0,002 |
| Within Groups | 625.351 | 23 | 27.189 |  |  |
| Total | 1062.740 | 25 |  |  |  |

Table 4
Spatial Recall Performance of Missing Features ANOVA Table

|  | Sum of Squares | df | Mean Square | F | Sig. |
|---|---|---|---|---|---|
| Between Groups | 103.349 | 2 | 51.675 | 7.170 | 0,004 |
| Within Groups | 165.766 | 23 | 7.170 |  |  |
| Total | 269.115 | 25 |  |  |  |

Table 5
Spatial Recall Performance of Misplaced Features ANOVA Table

|  | Sum of Squares | df | Mean Square | F | Sig. |
|---|---|---|---|---|---|
| Between Groups | 6.690 | 2 | 3.345 | 0.931 | 0,408 |
| Within Groups | 82.589 | 23 | 3.591 |  |  |
| Total | 89.279 | 25 |  |  |  |

Table 6
Spatial recall performance score averages of groups

| Groups | Feature Score (out of 30) | Missing Features Score | Misplaced Features Score |
|---|---|---|---|
| Real-World (N=7) | 7.43 | 10.86 | 2.57 |
| Video (N=8) | 9.5 | 9 | 3.88 |
| IVR (N=11) | 16.68 | 6.1 | 3.5 |

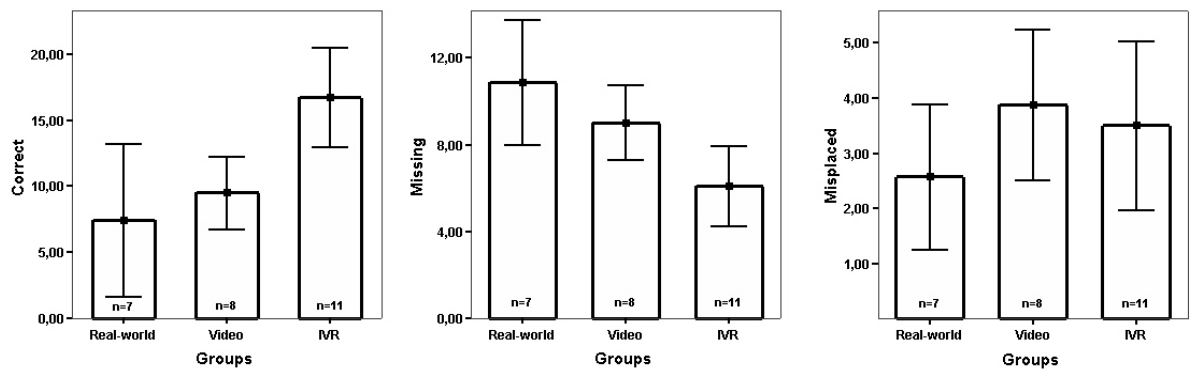

Fig 4
Summary graphs of findings (Error bars show 95% of CI of mean)

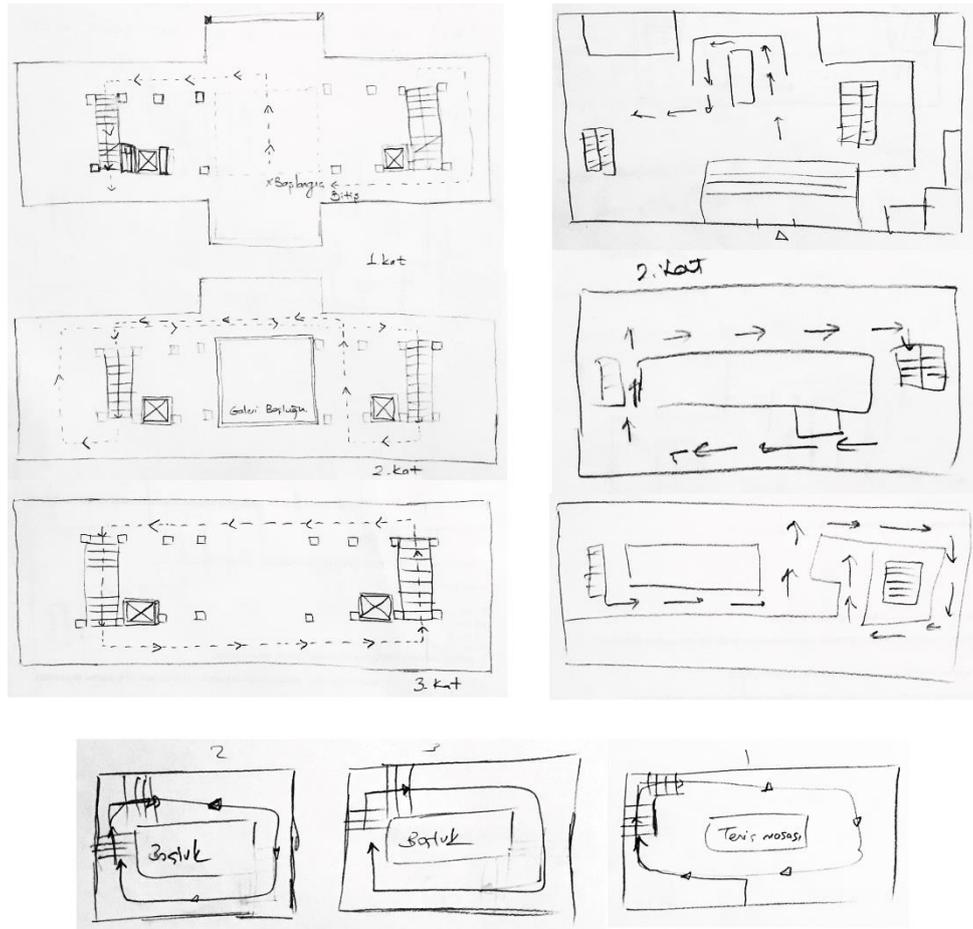

Fig 5
Example drawings of recall task.
Top left: Example from VR Group. Score: 22 out of 30; 4 missing; 3 errors. Top right: Example from Real-world Group. Score 18.5 out of 30. 5 missing; 1.5 errors. Bottom: Example from Video Group. Score: 9 out of 30; 10 missing; 2 errors.

## 5. Discussion

This study aimed to compare and investigate the efficacy of the direct-experiences, IVRE, and video representations on layout-learning. The hypothesis was that: (a) the students that experienced real-world and IVR would be able to recall more spatial memories than video walkthrough experiences; (b) real-world experiences will be superior to IVR experiences.

First part of the hypothesis, which is that the participants that experienced real world and IVR field trips, will be able to recall more spatial memories than video walkthrough experiences is partially accepted. The IVR group scored significantly more than both the video and real-world experience group. Second part of the hypothesis which is that the participants that experienced real-world will recall more spatial memories compared to IVR is rejected (Table 7).

As the theory explained before, the real-world provides the richest sensory input related to the environment, and IVR simulate this input, maybe not completely similar but almost similar at some level. Method – media debate of instructional technology also claims that media (IVR in this case) are only vehicles to transport the knowledge and will not influence learning (Clark, 1994), and the method is the main influencer of learning (field trip of a selected environment in this case).

Table 7
Summary of findings

|  | **IVR - Real-world** | **IVR - Video** | **Real-world - Video** |
|---|---|---|---|
| **Correct Features** | IVR (M = 16.68) > Real-world (M = 7.43)* | IVR (M = 16.68 > Video M = 9.50)* | No sig. differences<br>Real-world (M = 7.43), Video (M = 9.50) |
| **Missing Features** | IVR (M = 6.09) < Real-world (M = 10.86)* | No sig. differences<br>IVR (M = 6.09, Video M = 9.00) | No sig. differences<br>Real-world (M = 10.86), Video (M = 9.00) |
| **Misplaced Features** | No sig. differences<br>IVR (M = 3.50) > Real-world (M = 2.57) | No sig. differences<br>IVR (M = 3.50 > Video M = 3.88) | No sig. differences<br>Real-world (M = 3.50), Video (M = 3.88) |

*indicates significant difference at p>0.05 level

By means of the richness of the sensory input three conditions can be compared. For the real-world experience, the real-world surrounds the viewers' field of view, and as the viewer moves inside the environment, the surrounded view is updated depending on the viewpoint of the viewer. The viewer feels being inside the environment. Vestibular, muscular, and skeletal sensory systems support the consistency of the visual data with locomotion, angular transformation, and gyroscopic data. This creates a richer data flow from the sensory systems to the brain, thus may leading to requesting higher levels of attention.

The IVR equipment with tracked HMD's try to imitate these sensory input and the feeling of being inside the environment by creating surrounding stereoscopic visual data, and some degree of vestibular and muscular data originated from the (limited) locomotion of the user. Unfortunately, the simulation is not perfect as the locomotion is limited, and the field of view of the equipment is slightly narrower than the human eye, and the weight of the HMD exhausts the users rapidly.

Contrarily, a video presentation is a vista scale (Hegarty et al., 2006) representation and perceived as a series of images (Montello et. Al., 2004) instead of a surrounding environment and the viewer needs to convert these series of images to a three-dimensional mental representation of the environment to create the layout information of the space that demands more mental power compared with a surrounding environment.

So, it can be expected that the spatial recall performance of the real-world and IVR must be equal, or there must be a tendency to favour real-world experiences, while both are superior to video representations. But our findings lead to a different conclusion. Therefore, it is better to investigate the factors that can affect the spatial learning negatively. In other words, the factors that can decrease the level of spatial learning.

Our findings show participants of the real-world experience scored the worst. There are two possible explanations for this. The first one is, students were distracted because of the crowded group and environment. The design of this study tries to imitate the methods of architectural education. In architectural education, field trips are organized with the attendance of a class with mostly about 10 to 50 students at the same time. So, in real-world experience part of this study, participants experienced the building within a group of their peers guided by the researchers. Contrarily, IVR and video groups experienced the environment by their selves in the laboratory. A crowded group of students can cause distractions. Some participants can lose their attention to space and prefer to chat with their friends. Some participants prefer to interest in irrelevant detail while the lecturer talks about the spatial organization. These distractions can interfere with the level of spatial learning in a real-world group. For the IVR group, participants were experiencing the environment by their selves, and they know they were monitored by the researcher during the experience. So possibly they were more concentrated on the environment. The video group paid extreme attention to the walkthrough because although researchers did not explain them the objective of the experiment, they know they were in an experiment, and this could put them in an alert position. The second possible explanation which is also related to the first one is, in real-world experiences, because of too much detail on in space and the surfaces of the environment such as walls, doors, columns etc. participants become overloaded and cannot pay enough attention to space itself.

The second explanation is the effect active wayfinding on spatial learning, which was also mentioned by Farrell et. Al. (2003), and Tüzün, & Özdinç (2016). When someone guides the participant group, participants are less likely to pay attention to the layout and route decisions, because a guide makes these decisions for them. Active wayfinding decisions force participants to make judgments about the spatial relations of the building and this leads to a better learning. In this study participants of the IVR and video group experienced the environment all by themselves in a quiet laboratory. The video and real-world group did not make any wayfinding decisions.

Video group were both physically and psychologically passive. IVR group were guided by the signs in the IVR environment, and they were free to look around when they want to. They were physically active but psychologically almost passive. The real-world group followed the researcher guiding them on the route, they did not try to understand the spatial relations of the environment. They were just following the crowd. So, they were physically active but psychologically passive.

All these findings show, among these three groups the most active group (both physically and psychologically) is the IVR group. IVR group is also the less distracted group among the three groups. The outcome of this finding is that, although theoretically provides richer spatial data, field trips, because of distractions, are probably not that efficient to teach about the layout knowledge. More studies needed to understand and measure the effects of different group sizes, lectures during experiences, and efficiencies of single and group experiences.

## 6. Conclusion

In this study, the spatial knowledge acquisition of direct-experiences, IVR, and video representations were studied. This study has three main findings. First, the study shows layout learning tends to be overly sensitive to distractions. Lower levels of distraction may lead to a better levels of layout learning. Furthermore, the distractions caused by crowded groups may decrease the layout knowledge acquired by the students. More studies are needed to articulate the application of spatial learning with different group sizes to clearly understand the effect of group sizes in spatial learning. Second, the study shows, for the domain of layout knowledge learning, a remarkably simple IVR model is enough (as we used a very simple model). The three-dimensional models which can be used for educational IVR experiences do not need to be photorealistic and full of every detail. Further studies needed to clarify the effect of detail level of the three-dimensional model on layout learning. Third, although real-world experiences give direct information about the spaces, layout knowledge acquisition from the surrounding environment (real-world and stereoscopic VR) requires psychologically active wayfinding decisions. Lastly, this study does not cover the effect of the instruction about the spatial relations of the studied space during participants are experiencing the space. The effect of this kind of instruction on spatial learning must be further studied.

## References

Adithya, B., Kumar, B. P., Lee, H., Kim, J. Y., Moon, J. C., & Chai, Y. H. (2018). An experimental study on relationship between foveal range and FoV of a human eye using eye tracking devices. In *2018 International Conference on Electronics, Information, and Communication (ICEIC)* (pp. 1-5). IEEE.

Albert, W. S., Rensink, R. A., & Beusmans, J. M. (1999). Learning relative directions between landmarks in a desktop virtual environment. *Spatial cognition and computation*, *1*(2), 131-144.

Berkman, M. I., & Akan, E. (2018). *Presence and Immersion in Virtual Reality. Encyclopedia of Computer Graphics and Games, 1–10.* doi:10.1007/978-3-319-08234-9_162-1

Carassa, A., Geminiani, G., Morganti, F., & Varotto, D. (2002). Active and passive spatial learning in a complex virtual environment: The effect of efficient exploration. *Cognitive processing*, *3*(4), 65-81.

Cherney, I. D. (2008). Mom, Let Me Play More Computer Games: They Improve My Mental Rotation Skills. Sex Roles. 59, 776-786.

Clark, R. E. (1994). Media will never influence learning. *Educational technology research and development*, *42*(2), 21-29.

Clawson, D. M., Miller, M. S., Knott, B. A., & Sebrechts, M. M. (1998). Navigational training in virtual and real buildings. In *Proceedings of the Human Factors and Ergonomics Society Annual Meeting* (Vol. 42, No. 20, pp. 1427-1431). Sage CA: Los Angeles, CA: SAGE Publications.

Farrell, M. J., Arnold, P., Pettifer, S., Adams, J., Graham, T., & MacManamon, M. (2003). Transfer of route learning from virtual to real environments. *Journal of Experimental Psychology: Applied*, *9*(4), 219.

Hegarty, M., Montello, D. R., Richardson, A. E., Ishikawa, T., & Lovelace, K. (2006). Spatial abilities at different scales: Individual differences in aptitude-test performance and spatial-layout learning. *Intelligence*, *34*(2), 151-176.

Henry, D., & Furness, T. (1993). Spatial perception in virtual environments: Evaluating an architectural application. In *Proceedings of IEEE Virtual Reality Annual International Symposium* (pp. 33-40). IEEE.

Hirtle, S. (2009). Cognitive maps. In *Handbook of Research on Geoinformatics* (pp. 58-64). IGI Global.

Hölscher, C., Meilinger, T., Vrachliotis, G., Brösamle, M., & Knauff, M. (2006). Up the down staircase: Wayfinding strategies in multi-level buildings. *Journal of Environmental Psychology*, *26*(4), 284-299.

Johnson, D. M., & Stewart, J. E. (1999). Use of virtual environments for the acquisition of spatial knowledge: Comparison among different visual displays. *Military Psychology*, *11*(2), 129-148.

Lawton, C. A. (1996). Strategies for indoor wayfinding: The role of orientation. Journal of environmental psychology, 16(2), 137-145.

Miller, M. S., Clawson, D. M., & Sebrechts, M. M. (1999). Long-term retention of spatial knowledge acquired in virtual reality. In *Proceedings of the Human Factors and Ergonomics Society Annual Meeting* (Vol. 43, No. 22, pp. 1243-1246). Sage CA: Los Angeles, CA: SAGE Publications.

Montello, D. R., Waller, D., Hegarty, M., & Richardson, A. E. (2004). Spatial memory of real environments, virtual environments, and maps. *Human spatial memory: Remembering where*, 251-285.

Moeser, S. D. (1988). Cognitive mapping in a complex building. Environment and Behavior, 20(1), 21-49.

Münzer, S., & Stahl, C. (2011). Learning routes from visualizations for indoor wayfinding: Presentation modes and individual differences. *Spatial Cognition & Computation*, *11*(4), 281-312.

Richardson, A. E., Montello, D. R., & Hegarty, M. (1999). Spatial knowledge acquisition from maps and from navigation in real and virtual environments. *Memory & cognition*, *27*(4), 741-750.

Rossano, M. J., West, S. O., Robertson, T. J., Wayne, M. C., & Chase, R. B. (1999). The acquisition of route and survey knowledge from computer models. *Journal of Environmental Psychology*, *19*(2), 101-115.

Ruddle, R. A., Payne, S. J., & Jones, D. M. (1997). Navigating buildings in" desk-top" virtual environments: Experimental investigations using extended navigational experience. Journal of Experimental Psychology: Applied, 3(2), 143.

Slater, M. (2003). A note on presence terminology. *Presence connect*, *3*(3), 1-5.

Sorby, S. A. (2009). Educational research in developing 3-D spatial skills for engineering students. *International Journal of Science Education*, *31*(3), 459-480.

Terlecki, M. S., Newcombe, N. S. (2005). How Important Is the Digital Divide? The Relation of Computer and Videogame Usage to Gender Differences in Mental Rotation Ability. Sex Roles. 53/5-6, 433-441.

Tüzün, H., & Özdinç, F. (2016). The effects of 3D multi-user virtual environments on freshmen university students' conceptual and spatial learning and presence in departmental orientation. Computers & Education, 94, 228-240.

Thorndyke, P. W., & Hayes-Roth, B. (1982). Differences in spatial knowledge acquired from maps and navigation. *Cognitive psychology*, *14*(4), 560-589.

Uttal, D. H., Meadow, N. G., Tipton, E., Hand, L. L., Alden, A. R., Warren, C., & Newcombe, N. S. (2013). The malleability of spatial skills: A meta-analysis of training studies.*Psychological bulletin*,*139*(2), 352.

Waller, D., Hunt, E., & Knapp, D. (1998). The transfer of spatial knowledge in virtual environment training. *Presence*, *7*(2), 129-143.

Wallet, G., Sauzéon, H., Rodrigues, J., & N'Kaoua, B. (2008). Use of virtual reality for spatial knowledge transfer: Effects of passive/active exploration mode in simple and complex routes for three different recall tasks. In *Proceedings of the 2008 ACM symposium on Virtual reality software and technology* (pp. 175-178).

Wilson, P. N., Foreman, N., Gillett, R., & Stanton, D. (1997). Active versus passive processing of spatial information in a computer-simulated environment. *Ecological Psychology*, *9*(3), 207-222.